\documentclass[preprint,aps]{revtex4-1}
\usepackage[fleqn]{amsmath}
\usepackage{graphicx}
\usepackage{hyperref}
\usepackage{cleveref}
\usepackage{epstopdf}
\usepackage{slashed}
\usepackage{rotating}
\usepackage{float}
\newcommand{\be}{\begin{equation}}
\newcommand{\ee}{\end{equation}}
\newcommand{\bea}{\begin{eqnarray}}
\newcommand{\eea}{\end{eqnarray}}
\newcommand{\ba}{\begin{array}}
\newcommand{\ea}{\end{array}}

\begin{document}
\title{ Electromagnetic and Axial-Vector Form Factors of the Quarks and Nucleon}
\author{Harleen Dahiya}
\affiliation{Department of Physics,\\ Dr. B.R. Ambedkar National
Institute of Technology,\\ Jalandhar, 144011, India}
\author{Monika Randhawa}
\affiliation{University Institute of Engineering and Technology,
Panjab University, Chandigarh, 160014, India}

\begin{abstract}

In light of the improved precision of the experimental measurements and enormous theoretical progress, the nucleon form factors have been evaluated with an aim to understand how the static properties and dynamical behavior of nucleons emerge from the theory of strong interactions between quarks. We have analysed the  vector and axial-vector nucleon form factors ($G^{p,n}_{E,M}(Q^2)$ and $G^{p,n}_{A}(Q^2)$) using the spin observables in the chiral constituent quark model ($\chi$CQM) which has made a significant contribution to the unraveling of the internal structure of the nucleon in the nonperturbative regime. We have also presented a comprehensive analysis of the flavor decomposition of the form factors ($G^{q}_{E}(Q^2)$, $G^{q}_{M}(Q^2)$ and $G^{q}_{A}(Q^2)$for $q=u,d,s$) within the framework of $\chi$CQM with emphasis on the extraction of the strangeness form factors which are fundamental to determine the spin structure and test the chiral symmetry breaking effects in the nucleon. The $Q^2$ dependence  of the vector and axial-vector form factors of the nucleon has been studied using the conventional dipole form of parametrization.
The results are in agreement with the available experimental data.

\end{abstract}


\maketitle

\section{Introduction}

The flavor and spin structure of the nucleon play an essential role in understanding the dynamics of the theory of the strong interaction quantum chromodynamics (QCD). Even after extensive studies, confinement has limited our knowledge and the understanding of hadron internal structure continues to remain a major unresolved problem in high energy spin physics. Ever since  the  deep inelastic scattering (DIS) experiments discovered the composite nature of the proton, several interesting studies have been carried out experimentally and theoretically to understand the constituents of the nucleon \cite{point-like1,point-like2}. Even though the DIS with polarized beams and/or targets probe the spin carried by the quarks in the nucleon \cite{emc1,emc2,smc1,smc2,adams1,adams2,adams3,adams4,hermes_spin1,hermes_spin2}, the fundamental question of very small spin (only about 30\%) carried by the constituent quarks still remains to be the subject of much controversy.  The spin contribution of the strange quarks in the nucleon is a nontrivial aspect and is of intense theoretical interest because of model assumptions and experimental limitations. Further, the results revealed in the famous DIS experiments by the New Muon Collaboration (NMC) \cite{nmc1,nmc2}, Fermilab E866 \cite{e8661,e8662,e8663}, Drell-Yan cross section ratios of the NA51 experiments \cite{baldit} and HERMES \cite{hermes_flavor} have revealed the presence of sea quarks indicating more subtle dynamics which should be nonperturbative in nature.

The electromagnetic form factors data has been obtained from the cross sections
data using the Rosenbluth separation method  \cite{RSM1,RSM2,RSM3,RSM4,RSM5,RSM6,RSM7,RSM8,RSM9}, the double polarization experiments by the Jefferson Lab (JLab) \cite{jlab1,jlab2,jlab3,jlab4} as well as from the Continuous Electron Beam Accelerator Facility (CEBAF) at JLab \cite{CEBAF-proton1,CEBAF-proton2,CEBAF-neutron1,CEBAF-neutron2,CEBAF-neutron3}.  Recent experiments at JLab have increased the $Q^2$ range of the form factors \cite{jlab5,jlab6,jlab7} and have triggered much
activity in the determination of the quark flavor contributions to the form
factors of the nucleon.  Further, during the last few years, the standard electroweak theory has provided a firm basis for the role of weak interaction as a precision probe of the nucleon structure. The study of the electromagnetic structure of the nucleon involves the vector electric and magnetic form factors $G^{\gamma,N}_E(Q^2)$ and $G^{\gamma,N}_M(Q^2)$ whereas the neutral weak interaction between leptons and nucleons involves vector weak form factors $G^{Z,N}_E(Q^2)$ and $G^{Z,N}_M(Q^2)$ as well as axial form factor $G^{Z,N}_A(Q^2)$.

The contribution of strange quarks to the nucleon structure is of special interest because it provides an ideal probe for the virtual sea quarks present in the nucleon. The strange spin polarization  $\Delta s$ has received much attention in the past as it corresponds to the value of the strange axial form factor $G_A^s$ at zero-momentum transfer ($Q^2 = 0$).  Over the last decade, several experiments have been proposed to probe the electromagnetic and the weak structure of the nucleon. There is a large effort to look for contribution of the sea quarks in the vector form factors via the parity-violating (PV) electron scattering providing information on the weak structure of the nucleon and their associated quark structure. PV electron scattering measurements which are sensitive to the strange quark contributions but not to the axial-vector form factor have been carried out by various collaborations \cite{sample,happex1,happex2,happex3,happex4,g01,g02,a41,a42,e1581,e1582}.  The DIS of neutrinos or of polarized charged leptons  from nucleon and nuclear targets have been used to measure the electromagnetic and axial form factors of the nucleon in the elastic $\nu p$ and $\bar{\nu} p$ scattering from the BNL E734 experiment \cite{e734},  Fermilab Intense Neutrino Scattering Scintillator Experiment (FINeSSE) \cite{finesse} at Fermi National Accelerator Laboratory (Fermilab). A determination of the strange form factors through a combined analysis of elastic $\nu p$ and $\bar{\nu} p$ and PV electron scattering is performed in Ref. \cite{pate}.

Even though many experimental and theoretical efforts have been made to understand the internal structure of the hadrons and origin of the sea quarks \cite{ellis-brodsky,alkofer,christov1,christov2,diakonov1,diakonov2,mesoncloud1,mesoncloud2,mesoncloud3,mesoncloud4,mesoncloud5,mesoncloud6,mesoncloud7,wakamatsu1,wakamatsu2,wakamatsu3,wakamatsu4,wakamatsu5,wakamatsu6,eccm1,eccm2,stat1,stat2,stat3,stat4,stat5,stat6,alwall,reya1,reya2,reya3,chang-141,chang-142}, the effective interaction Lagrangian approach of the strong interactions used in the chiral constituent quark model ($\chi$CQM) \cite{manohar1,manohar2,eichten,cheng1,cheng2,cheng3,johan1,johan2,song1,song2} successfully explains  the spin structure of the nucleon \cite{hd1,hd2,hd3,hd4,hd5,hd6,hd7,hd8,hd9,hd10,hd11}, magnetic moments of octet and decuplet baryons \cite{hdmagnetic1,hdmagnetic2}, semileptonic weak decay parameters \cite{nsweak1,nsweak2}, magnetic moments of nucleon resonances and $\Lambda$ resonances \cite{nres-torres1,nres-torres2}, quadrupole moment and charge radii of octet baryons \cite{charge-radii1,charge-radii2}, etc..  The inclusion of $Q^2$ dependence in the vector and axial-vector form factors can be done through the dipole form of parametrization as  \be G^{p,n}_{V,A}(Q^2)=\frac{g^{p,n}_{V,A}(0)}{\left( 1+\frac{Q^2}{M_{V,A}^2}\right)^2}, \label{dipole}\ee where $M_V$  and $M_A$  are the canonical vector and axial-vector masses respectively. The factors $g_V^{p,n}(0)$ are the charge, magnetic moment of the proton and neutron at zero momentum transfer whereas  $g_A^{p,n}(0)$ is the isovector axial-vector coupling constant of the proton and neutron at zero momentum transfer. In view of the above developments, it becomes desirable to extend the applicability of $\chi$CQM by incorporating $Q^2$ dependence phenomenologically in the proton and neutron form factors whose knowledge would undoubtedly provide vital clues to the nonperturbative aspects of QCD.

\section{Dirac and Pauli Nucleon Form Factors}

The most general form for the hadronic matrix element of the electromagnetic current operators for a spin$-\frac{1}{2}$ nucleon  with
internal structure, in terms of the Dirac and Pauli form factors $F^{N,\gamma}_1$ and $F^{N,\gamma}_2$ ($N=p,n$),  satisfying relativistic invariance and current conservation, is expressed as
\be
J _{\mu}^{EM}= e \bar{U}(p′) \left[ \gamma^{\mu} F^{N,\gamma}_1(Q^2) + \frac{i \sigma^{\mu \nu} q_{\nu}}{2M} F^{N,\gamma}_2(Q^2) \right] U(p),
\ee
where $M$
is the nucleon mass, $Q^2=−q^2$ is the negative of the square of the invariant mass of the virtual photon in the one-photon exchange
approximation in $ep$ scattering. In the static limit $Q^2 = 0$, the form factors give the charge and anomalous magnetic moment as  $F_1^{p,\gamma}(0) = 1$, $F_2^{p,\gamma}(0) = \kappa^p$, $F_1^{n,\gamma}(0) = 0$, $F_2^{n,\gamma}(0) =
\kappa ^n$, of the proton and neutron respectively. The anomalous  magnetic moment and the magnetic moment of the proton and neutron are related as $\kappa^p = \mu^{p}-1$ and  $\kappa^n = \mu^n$ respectively.

The Dirac and Pauli form factors are related to the electric and magnetic Sachs form factors as
\bea
G^{N,\gamma}_E(Q^2)&=&F^{N,\gamma}_1(Q^2)-\tau F^{N,\gamma}_2(Q^2)\,,\nonumber \\
G^{N,\gamma}_M(Q^2)&=&F^{N,\gamma}_1(Q^2)+ F^{N,\gamma}_2(Q^2)\,, \label{gEM-F12}
\eea
where $\tau=\frac{Q^2}{4M^2}$. In the static limit $Q^2 = 0$, the electric and magnetic form factors give the charge and magnetic moments of the proton
and neutron, respectively as $G_E^{p,\gamma}(0) = 1$, $G_M^{p,\gamma}(0) = \mu^p$, $G_E^{n,\gamma}(0) = 0$, $G_M^{n,\gamma}(0) =
\mu ^n$.

The quark flavor structure of these form factors can be revealed from the
matrix elements of individual quark currents in terms of form factors $F^{q}_1$ and $F^{q}_2$ ($j=u$, $d$, or $s$)
\be
J _{\mu}^{q}= e \bar{U}(p′) \left[ \gamma^{\mu} F^{q}_1(Q^2) + \frac{i \sigma^{\mu \nu} q_{\nu}}{2M} F^{q}_2(Q^2) \right] U(p)\,.
\ee
Because of the point-like interaction between electrons and the
quark constituents of the nucleon, these nucleon form factors can be expressed in terms of the individual quark flavor contributions with the electric charge  of individual quarks as the coupling constants. These quark flavor contributions to the form factors are then global properties of the nucleons. Using the definitions analogous to Eq. (\ref{gEM-F12}), we can write
\bea
G^{p,\gamma}_{E,M}(Q^2)&=&\frac{2}{3} G_{E,M}^{u}(Q^2)-\frac{1}{3} G_{E,M}^{d}(Q^2)-\frac{1}{3} G_{E,M}^{s}(Q^2)\,,\nonumber \\
G^{n,\gamma}_{E,M}(Q^2)&=&\frac{2}{3} G_{E,M}^{d}(Q^2)-\frac{1}{3} G_{E,M}^{u}(Q^2)-\frac{1}{3} G_{E,M}^{s}(Q^2)\,.
\eea
Here we have assumed charge symmetry for the rotation transformation of $\frac{\pi}{2}$ in the isospin space between $p \leftrightarrow n$.
The strange form factors in each nucleon are also taken to be the same. This can be used to calculate the flavor decomposition of the form factors by using the well known electromagnetic form factors of the proton and neutron at low $Q^2$. The calculation of the strangeness form factors $G_E^{s}(Q^2)$ and $G_M^{s}(Q^2)$ however requires the information from the neutral weak current.

The hadronic matrix element of the neutral weak current operators for a spin$-\frac{1}{2}$ nucleon can be expressed in terms of the vector form factors $F^{N,Z}_1$ and $F^{N,Z}_2$ as well as the axial form factor $G^{N,Z}_A$ as
\be
J _{\mu}^{NC}= e \bar{U}(p′) \left[ \gamma^{\mu} F^{N,Z}_1(Q^2) + \frac{i \sigma^{\mu \nu} q_{\nu}}{2M} F^{N,Z}_2(Q^2)+\gamma^{\mu} \gamma_5 G^{N,Z}_A(Q^2)  \right] U(p).
\ee
The nucleon form factors in terms of the quark flavor contributions to the form
factors with the weak electric charge $e^Z$ of individual quarks ($e^Z=\left (1-\frac{8}{3} \sin^2 \theta_W \right )$ for $u$ and $e^Z=\left(-1+\frac{4}{3} \sin^2 \theta_W \right)$ for $d$ and $s$ quarks) as the coupling constants and the weak mixing angle $\theta_W$ can be expressed as
\be
G^{p,Z}_{E,M}(Q^2)=\left (1-\frac{8}{3} \sin^2 \theta_W \right )G_{E,M}^{u}(Q^2)+\left(-1+\frac{4}{3} \sin^2 \theta_W \right) \left( G_{E,M}^{d}(Q^2)+ G_{E,M}^{s}(Q^2)\right)\,,\ee
\be
G^{n,Z}_{E,M}(Q^2)=\left (1-\frac{8}{3} \sin^2 \theta_W \right )G_{E,M}^{d}(Q^2)+\left(-1+\frac{4}{3} \sin^2 \theta_W \right ) \left(G_{E,M}^{u}(Q^2)+ G_{E,M}^{s}(Q^2)\right )\,.
\ee
Utilizing the isospin symmetry at leading order as well as the proton and neutron
electromagnetic form factors, the up and down quark
contributions to the neutral weak form factors can be eliminated to obtain the contribution of strange quarks as \cite{beck-mckeown2001} 
\be
G_{E,M}^s(Q^2)= \left(1-4 \sin^2 \theta_W \right )G_{E,M}^{p,\gamma}(Q^2)-G_{E,M}^{n,\gamma}(Q^2)-G_{E,M}^{p,Z}(Q^2)\,.
\ee
This clearly shows how the contribution from the strange form factor is related to the electromagnetic form factors  as well as the neutral weak form factors. Therefore, the  measurement of the neutral weak form factor, in combination with the electromagnetic form factors, will
allow the determination of the strange form factor.

\section{Nucleon Form Factors in the Chiral Constituent Quark Model ($\chi$CQM)}

Since the presence of sea quarks is a nonperturbative in nature, the $\chi$CQM uses the effective interaction Lagrangian approach where the chiral symmetry breaking takes place at a distance scale much smaller than the confinement scale. The dynamics of light quarks ($u$, $d$, and $s$) and gluons is described by  the  QCD Lagrangian as
\be
{\cal{L}} = i \bar{\psi}_L \slashed{D}
{\psi}_L + i
\bar{\psi}_R \slashed{D} {\psi}_R  - \bar{\psi}_L M {\psi}_R - \bar{\psi}_R M {\psi}_L - \frac{1}{4}G_{\mu \nu}^{a} G^{\mu \nu}_{a} \,,
\label{lagrang1} \ee where $\psi_L$  and $\psi_R$ are the left and right
handed quark fields respectively, $M$ is the quark mass matrix, $ G_{\mu \nu}^{a}$ is the gluonic gauge field strength tensor, and $D^{\mu}$ is the gauge-covariant derivative.
Under the chiral transformation $(\psi \to \gamma^5 \psi)$, the mass
terms change sign  as $\psi_{L}
\to -\psi_{L}$ and $\psi_{R} \to \psi_{R}$ and the Lagrangian in Eq. (\ref{lagrang1}) no longer remains invariant. In case the mass terms are neglected, the Lagrangian will have global chiral symmetry of
the {\it SU}(3)$_L$$\times${\it SU}(3)$_R$ group. The chiral symmetry is believed to be spontaneously broken to ${\it SU}(3)_{L+R}$ around the scale of 1\,GeV and as a consequence, a set of massless particles (referred to as the Goldstone
bosons (GBs)) exist. These GBs are identified with the observed ($\pi$, $K$,
$\eta$ mesons). Within the region of chiral symmetry breaking scale
$\Lambda_{\chi SB}$ and the QCD confinement scale
($\Lambda_{QCD} \simeq 0.1-0.3$\,GeV), the appropriate degrees of freedom are the constituent quarks, the set of GBs
($\pi$, K, $\eta$ mesons), and the {\it weakly} interacting gluons.

The effective interaction Lagrangian  between
GBs and quarks in the leading order can now be expressed as \be {\cal
	L}_{{\rm int}} = -\frac{g_{A}}{f_{\pi}} \bar{\psi} \partial_{\mu}
\Phi \gamma^{\mu} \gamma^{5} \psi \,, \label{lagrang3} \ee where the field $\Phi$ describes the dynamics
of octet of GBs.
The QCD Lagrangian is also invariant under the axial
$U(1)$ symmetry, which would imply the existence of ninth GB. This
breaking symmetry picks the $\eta'$ as the ninth GB. The effective
Lagrangian describing interaction between quarks and a nonet of GBs,
consisting of octet and a singlet, can now be expressed as \be {\cal
	L}_{{\rm int}} = g_8 { \bar \psi} \Phi {\psi} + g_1{ \bar \psi}
\frac{\eta'}{\sqrt 3}{\psi}= g_8 {\bar \psi}\left( \Phi + \zeta
\frac{\eta'}{\sqrt 3}I \right) {\psi }=g_8 {\bar \psi} \left(\Phi'
\right) {\psi} \,, \label{lagrang4} \ee where $\zeta=g_1/g_8$, $g_1$ ($g_8$)
is the coupling constant for the singlet (octet) GB and $I$ is the $3\times 3$
identity matrix.

The basic idea in the $\chi$CQM \cite{manohar1,manohar2} is the fluctuation process where the
GBs are emitted by a constituent quark. These GBs further split into a $q
\bar q$ pairs, for example,
\be q^{\uparrow(\downarrow)} \rightarrow {\rm GB}^0 + q^{'
	\downarrow(\uparrow)} \rightarrow (q \bar q^{'})^0 +q^{'\downarrow(\uparrow)}\,, \label{basic}
\ee
where $q \bar q^{'} +q^{'}$
constitute the sea quarks \cite{cheng1,cheng2,cheng3,johan1,johan2,hd1,hd2,hd3,hd4,hd5,hd6,hd7,hd8,hd9,hd10,hd11}. The GB field can be expressed in terms of the GBs and their transition probabilities as \bea
\Phi' = \left( \ba{ccc} \frac{\pi^0}{\sqrt 2}
+\beta\frac{\eta}{\sqrt 6}+\zeta\frac{\eta^{'}}{\sqrt 3} & \pi^+
& \alpha K^+   \\
\pi^- & -\frac{\pi^0}{\sqrt 2} +\beta \frac{\eta}{\sqrt 6}
+\zeta\frac{\eta^{'}}{\sqrt 3}  &  \alpha K^o  \\
\alpha K^-  &  \alpha \bar{K}^0  &  -\beta \frac{2\eta}{\sqrt 6}
+\zeta\frac{\eta^{'}}{\sqrt 3} \ea \right). \eea
The transition probability of chiral
fluctuation  $u(d) \rightarrow d(u) + \pi^{+(-)}$, given in terms of the coupling constant for the octet GBs $|g_8|^2$, is defined as $a$ and is introduced by considering nondegenerate quark masses $M_s > M_{u,d}$. The probabilities of transitions of $u(d) \rightarrow s + K^{+(0)}$, $u(d,s)\rightarrow u(d,s) + \eta$, and $u(d,s) \rightarrow u(d,s) + \eta^{'}$ are given as $\alpha^2 a$, $\beta^2 a$ and $\zeta^2 a$ respectively \cite{cheng1,cheng2,cheng3,johan1,johan2}. The probability parameters $\alpha^2 a$ and $\beta^2 a$ are introduced by considering nondegenerate GB masses $M_{K},M_{\eta}> M_{\pi}$ and the probability $\zeta^2 a$ is introduced by considering  $M_{\eta^{'}} > M_{K},M_{\eta}$.

The calculations of vector and axial-vector form factors involve the calculations of  axial-vector matrix elements of the nucleons using the operator $q^{\uparrow} q^{\downarrow}$ measuring the sum of the quark with spin up and down as
\be
\langle p(n))|q^{\uparrow}q^{\downarrow}|p(n)\rangle\,. \label{BNB}
\ee
Here  $q^{\uparrow}q^{\downarrow}$ is the number
operator defined in terms of the number $n^{q^{\uparrow}(q^{\downarrow})}$ of $q^{\uparrow}({q^{\downarrow}})$ quarks and is expressed as
\be
q^{\uparrow}q^{\downarrow}=\sum_{q=u,d,s} (n^{q^{\uparrow}}q^{\uparrow} + n^{q^{\downarrow}}q^{\downarrow})=n^{u^{\uparrow}}u^{\uparrow} + n^{u^{\downarrow}}u^{\downarrow} + n^{d^{\uparrow}}d^{\uparrow} + n^{d^{\downarrow}}d^{\downarrow} +
n^{s^{\uparrow}}s^{\uparrow} + n^{s^{\downarrow}}s^{\downarrow}\,, \label{number2}
\ee
with the coefficients of the $q^{\uparrow\downarrow}$ giving the number of
$q^{\uparrow\downarrow}$ quarks.

The spin structure of the nucleon after the inclusion of sea quarks generated
through chiral fluctuation can be calculated by substituting for
each constituent quark \be q^{\uparrow\downarrow} \rightarrow P^q q^{\uparrow\downarrow}+
|\psi^{q^{\uparrow\downarrow}}|^2 \,, \label{spin} \ee where the transition probability of no emission of
GB $P^q$ can be expressed in terms of the transition probability of the emission of a GB from
any of the $u$, $d$, and $s$ quark as follows
\be
P^q=1-P^{[q, ~GB]}, \label{probability} \ee with \be P^{[u, ~GB]}=P^{[d, ~GB]}
=\frac{a}{6}\left(9+6\alpha^2+\beta^2+2\zeta^2\right)\,,~~~~{\rm and}~~~~
P^{[s, ~GB]} = \frac{a}{3}\left(6 \alpha^2+2\beta^2+\zeta^2 \right)\,. \label{probuds} \ee
The probabilities of
transforming $q^{\uparrow\downarrow}$ quark after one interaction $|\psi^{q^{\uparrow\downarrow}}|^2$ are expressed by the
functions \bea |\psi^{u^{\uparrow\downarrow}}|^2 &=& \frac{a}{6}\left(3 +
\beta^2 + 2 \zeta^2 \right)u^{\downarrow\uparrow}+ a d^{\downarrow\uparrow} + a \alpha^2
s^{\downarrow\uparrow}\,, \nonumber \\ |\psi^{d^{\uparrow\downarrow}}|^2
&=& a u^{\downarrow\uparrow}+ \frac{a}{6} \left(3+\beta^2+2 \zeta^2 \right)d^{\downarrow\uparrow}+
a \alpha^2 s^{\downarrow\uparrow}\,, \nonumber\\ |\psi^{s^{\uparrow\downarrow}}|^2 &=& a \alpha^2 u^{\downarrow\uparrow} + a\alpha^2 d^{\downarrow\uparrow} +
\frac{a}{3} \left(2 \beta^2 + \zeta^2 \right)s^{\downarrow\uparrow} \,.
\label{psiup} \eea

The vector and axial-vector form factors of the nucleon at $Q^2=0$ are given  by the polarized distribution function of the quark $\Delta q$ in the $\chi$CQM defined as
\be
\Delta q= q^{\uparrow}- q^{\downarrow}, \label{deltaq}
\ee
where $q^{\uparrow}$ $(q^{\downarrow})$ is the probability that the quark spin is aligned parallel or antiparallel to the nucleon spin. This can further
be defined in terms of polarized constituent (C) and sea (S) quark distribution functions as
\be \Delta q^{p,n}=\Delta q^{p,n}_{{\rm C}}+\Delta q^{p,n}_{{\rm S}}\,. \label{totalquarksea}\ee
Here we have the polarized constituent quark distribution functions for $p$ and $n$ as
\bea
\Delta u^p_{{\rm C}}=\frac{4}{3}, &~~~& \Delta d^p_{{\rm C}}=-\frac{1}{3},~~~ \Delta s^p_{{\rm C}}=0\,, \nonumber \\
\Delta u^n_{{\rm C}}=-\frac{1}{3}, &~~~& \Delta d^n_{{\rm C}}=\frac{4}{3}, ~~~\Delta s^n_{{\rm C}}=0\,, \label{deltaqval}
\eea
and the polarized sea quark distribution functions for $p$ and $n$ as
\bea
\Delta u^p_{{\rm S}}= -\frac{a}{3} (7+4 \alpha^2+
\frac{4}{3}\beta^2 +\frac{8}{3} \zeta^2)\,, &~~~& \Delta u^n_{{\rm S}}=-\frac{a}{3} (2-\alpha^2
-\frac{1}{3}\beta^2 -\frac{2}{3} \zeta^2)\,,  \nonumber \\
\Delta d^p_{{\rm S}}=-\frac{a}{3} (2-\alpha^2
-\frac{1}{3}\beta^2 -\frac{2}{3} \zeta^2)\,, &~~~& \Delta d^n_{{\rm S}}= -\frac{a}{3} (7+4 \alpha^2+
\frac{4}{3}\beta^2 +\frac{8}{3} \zeta^2)\,,\nonumber \\
\Delta s^p_{{\rm S}}= -a \alpha^2\,, &~~~& \Delta s^n_{{\rm S}} = -a \alpha^2\,. \label{deltaqsea}
\eea

In terms of the polarized distribution functions, the magnetic moment of the nucleon is defined as
\be
\mu^{p,n}=\sum_{q=u,d,s}{\Delta q^{p,n}\mu^q}.
\ee
Apart from the spin of the the constituent quarks and spin of the sea quarks, the magnetic moment of a given baryon in the $\chi$CQM also receives
contribution from the orbital angular motion of the sea quarks. The total magnetic moment is expressed as \be \mu^{p,n}= \mu_{{\rm C}}^{p,n}+\mu_{{\rm S}}^{p,n} + \mu_{{\rm O}}^{p,n}\,, \label{totalmag} \ee
where $\mu_{{\rm C}}^{p,n}$ and $\mu_{{\rm S}}^{p,n}$ are the  magnetic moment contributions of the constituent quarks and the sea quarks respectively coming from the proton and neutron spin polarizations, whereas  $\mu_{{\rm O}}^{p,n}$ is the magnetic moment contribution due to the rotational
motion of the two bodies constituting the sea quarks ($q^{'}$) and GB and referred to as the orbital
angular momentum contribution of the quark sea \cite{cheng1,cheng2,cheng3}.

In terms of quark magnetic moments and spin polarizations, the contributions of
constituent quark spin ($\mu^{p,n}_{{\rm C}}$), sea quark spin ($\mu^{p,n}_{{\rm S}}$), and sea orbital ($\mu^{p,n}_{{\rm O}}$) can be defined as
\bea
\mu_{{\rm C}}^{p,n} &=& \sum_{q=u,d,s}{\Delta q^{p,n}_{{\rm C}}\mu^q}\,,\label{mag-val}\\
\mu_{{\rm S}}^{p,n} &=& \sum_{q=u,d,s} {\Delta q^{p,n}_{{\rm
			S}}\mu^q}\,, \label{mag-sea} \\ \mu_{{\rm O}}^{p,n} &=& \sum_{q=u,d,s} {\Delta
	q^{p,n}_{{\rm C}}~\mu(q_{+} \rightarrow )} \,,\label{mag-orbit}
\eea
where $\mu^q= \frac{e^q}{2 M_q}$ ($q=u,d,s$) is the quark magnetic moment in the units of $\mu^N$ (nuclear magneton), $e^q$ and $M_q$ are the electric charge and the mass,
respectively, for the quark $q$. $\Delta q^{p,n}_{{\rm C}}$ and  $\Delta q^{p,n}_{{\rm S}}$ can be calculated from Eqs. (\ref{deltaqval}) and (\ref{deltaqsea}). The orbital moment for any chiral
fluctuation
$\mu(q^{\uparrow} \rightarrow )$ can be calculated from the contribution of the
angular momentum of the sea quarks to the magnetic moment of a given
quark expressed as
\be
\mu (q^{\uparrow} \rightarrow {q}^{'\downarrow}) =\frac{e^{q^{'}}}{2M_q}
\langle l^q \rangle +
\frac{{e}^{q}-{e}^{q^{'}}}{2 {M}_{{\rm GB}}}\langle {l}^{{\rm GB}} \rangle\,,
\ee
where
\be
\langle l^q \rangle=\frac{{M}_{{\rm GB}}}{M_q+{M}_{{\rm GB}}} ~{\rm and}
~\langle l^{{\rm GB}} \rangle=\frac{M_q}{M_q+{M}_{{\rm GB}}}\,,
\ee
$\langle l^q, l^{{\rm GB}} \rangle$ and ($M_q$, ${M}_{{\rm GB}}$) are the
orbital angular momenta and masses of quark and GB respectively.
The orbital moment of each process is then multiplied by the probability
for such a process to take place to yield the magnetic moment due to
all the transitions starting with a given constituent quark, for example
\bea
[ \mu (u^{\uparrow\downarrow}(d^{\uparrow\downarrow}) \rightarrow )] &=&  \pm a
[\mu \left(u^{\uparrow}(d^{\uparrow}) \rightarrow d^\downarrow (u^\downarrow)\right) +
\alpha^2 \mu \left(u^\uparrow(d^\uparrow) \rightarrow s^\downarrow\right) \nonumber \\
&+& \left.\left(\frac{1}{2} +\frac{1}{6} \beta^2+ \frac{1}{3} \zeta^2\right)
\mu \left(u^{\uparrow}(d^{\uparrow}) \rightarrow u^\downarrow (d^\downarrow)\right)\right], \label{mud}
\eea
\be
[\mu (s^{\uparrow\downarrow} \rightarrow )] =  \pm a
\left[\alpha^2 \mu \left(s^{\uparrow} \rightarrow u^\downarrow\right) +
\alpha^2 \mu \left(s^\uparrow \rightarrow d^\downarrow\right) +
\left(\frac{2}{3} \beta^2+ \frac{1}{3} \zeta^2\right)
\mu \left(s^{\uparrow} \rightarrow s^\downarrow \right)\right]. \label{mus}
\ee
The above equations can easily be
generalized by including the coupling breaking and mass
breaking terms, for example, in terms of the coupling breaking
parameters $a$, $\alpha$, $\beta$ and $\zeta$  as well as the masses
of GBs $M_{\pi}$, $M_{K}$
and $M_{\eta}$.

\section{$Q^2$ Dependence of Nucleon Form Factors}

The $Q^2$ dependence of the vector electric and magnetic form factors as well as axial-vector form factors have been experimentally investigated from the PV electron scattering  and from the DIS of neutrinos. The conventional dipole form of parametrization has been used to analyse the vector and axial-vector form factors
\be
G^{p,n}_{V,A}(Q^2)=g^{p,n}_{V,A}(0)G^D_{V,A}(Q^2)= \frac{g^{p,n}_{V,A}(0)}{\left( 1+\frac{Q^2}{M_{V,A}^2}\right)^2},
\label{dipole}\ee
where  the electric and magnetic form factors of the proton and neutron  at zero momentum transfer $g^{p,n}_{V}(0)$ for $V=E,M$ correspond to the charge and magnetic moment respectively. $g_A^{p}(0)$ and $g_A^{n}(0)$  are the isovector axial-vector coupling constants of the proton and neutron corresponding to the axial-vector form factors at zero momentum transfer. The vector mass $M_V$ is taken as $M_V^2 = 0.71$\,GeV$^2$. For the axial mass $M_A$, we have used the most recent value obtained by the MiniBooNE Collaboration $M_A^2 = 1.10^{+0.13}_{-0.15}$\,GeV$^2$ \cite{miniboone1,miniboone2}.

For the case of proton, both the vector and axial-vector form factors of the proton follow the dipole form of parametrization. The form factors $G^{p}_{E}(Q^2)$, $G^{p}_{M}(Q^2)$ and $G^{p}_{A}(Q^2)$   respectively scale with the net charge, magnetic moment $\mu_p$ and isovector axial-vector coupling constant $g_A^{i}(0)$ ($i=0,3,8$) as follows
\bea
G^{p}_{E}(Q^2)&=& G^D_V(Q^2)=\frac{1}{\left( 1+\frac{Q^2}{M_{V}^2}\right)^2}\,, \nonumber \\
G^{p}_{M}(Q^2)&=& \mu_p G^D_V(Q^2)=\frac{\mu_p}{\left( 1+\frac{Q^2}{M_{V}^2}\right)^2}\,, \nonumber \\
G^{i}_{A}(Q^2)&=& g_A^{i}(0)G^D_A(Q^2)=\frac{g_A^{i}(0)}{\left( 1+\frac{Q^2}{M_{A}^2}\right)^2}\,.
\eea

For the case of neutron however, the measurement of form factors raises difficulties because there are no free neutron target available  suited for electron experiments.   The form factors $G^{n}_{E}(Q^2)$ and $G^{n}_{M}(Q^2)$   respectively scale with the net charge and magnetic moment  as follows
\bea
G^{n}_{E}(Q^2)&=& \frac{A\tau}{1+B\tau}G^D_V(Q^2)=\frac{A \tau}{1+B\tau}\frac{1}{\left( 1+\frac{Q^2}{M_{V}^2}\right)^2}\,, \nonumber \\
G^{n}_{M}(Q^2)&=& \mu_n G^D_V(Q^2)=\frac{\mu_n}{\left( 1+\frac{Q^2}{M_{V}^2}\right)^2}\,,
\eea
where $\tau=\frac{Q^2}{4 M_n^2}$ and the parameters $A$ and $B$ obtained from the recent fits of root mean square radius are $A = 1.73$ and $B= 4.62$. The factor $\frac{A \tau}{1+B \tau}$ has been introduced to basically account for the condition $G^{n}_{E}(Q^2=0)=0$.

\section{Quark Flavor Decomposition of the Form Factors}

Recent measurements of form factors have made it possible to separate out the quark flavor contributions
which have been subject of extensive
theoretical analysis and are not understood completely by existing models. The  quark flavor form factors at $Q^2=0$ can be calculated in the $\chi$CQM and are expressed as

\bea
G^{q}_{E}(Q^2=0)=e^q\,, \nonumber \\
G^{q}_{M}(Q^2=0)=\mu^q\,, \nonumber \\
G^{q}_{A}(Q^2=0)=\Delta q\,.
\eea
The $Q^2$ dependence of these form factors can be calculated using the  dipole parametrization in Eq. (\ref{dipole}).
We have the  quark flavor form factors corresponding to the nucleon as
\bea
G^{q}_{E}(Q^2)&=& e^q G^D_V(Q^2)=\frac{e^q}{\left( 1+\frac{Q^2}{M_{V}^2}\right)^2}\,, \nonumber \\
G^{q}_{M}(Q^2)&=& \mu_{p,n}^q G^D_V(Q^2)=\frac{\mu_{p,n}^q}{\left( 1+\frac{Q^2}{M_{V}^2}\right)^2}\,, \nonumber \\
G^{q}_{A}(Q^2)&=& \Delta q G^D_A(Q^2)=\frac{\Delta q}{\left( 1+\frac{Q^2}{M_{A}^2}\right)^2}\,.
\eea

Out of all the flavor vector and axial-vector form factors, the strangeness form factors have triggered a great deal of interest.
The recent measurements by  SAMPLE  at MIT-Bates \cite{sample}, HAPPEX
\cite{happex1,happex2,happex3,happex4}, G0  at JLab \cite{g01,g02}, A4  at MAMI \cite{a41,a42} have observed either one or a combination of electric and
magnetic form factors.  Recent lattice calculations \cite{lattice1,lattice2,lattice3} and other phenomenological studies with lattice
inputs \cite{mup-aw1,mup-aw2} have also predicted a very small value for the
strange contribution.

\section{results and discussion}
The probabilities of fluctuations to
pions, $K$, $\eta$, $\eta^{'}$ represented by $a$, $a \alpha^2$, $a \beta^2$, and $a \zeta^2$ respectively can be calculated in the $\chi$CQM at $Q^2=0$ after taking into account strong physical considerations and carrying out a fine grained analysis using the well known experimentally measurable spin and flavor distribution functions. The parameters are listed in Table \ref{input}. The table also includes the other input parameters pertaining to quark masses and magnetic moments as well as the GB masses.

\begin{table}
\begin{center}
\begin{tabular}{|c|c|c|c|c|c|c|c|c|c|c|c|c|c|c|c|}      \hline
Parameter&$a$ & $a \alpha^2$ &
$a \beta^2$ &
$a\zeta^2$ &
$M_{u,d}$ &
$M_s$ &
$\mu^u$ &
$\mu^d$ &
$\mu^s$ &
$M_{\pi}$ &
$M_{K}$ &
$M_{\eta}$ &
$M_{\eta'}$ &
$A$ &
$B$   \\ \hline

Value&  0.114
& 0.023  & 0.023 & 0.002  & $330$& $510$ & 2& -1 & -0.65 & 140  & 494  & 548  & 958  & $1.73$ & $4.62$ \\
 \hline
\end{tabular}
\end{center}
\caption{The numeric values of input parameters. The units of  masses are in MeV and the magnetic moments in $\mu_N$ (nuclear magneton). }
\label{input}
\end{table}


\begin{figure}
\includegraphics[width=1.\textwidth] {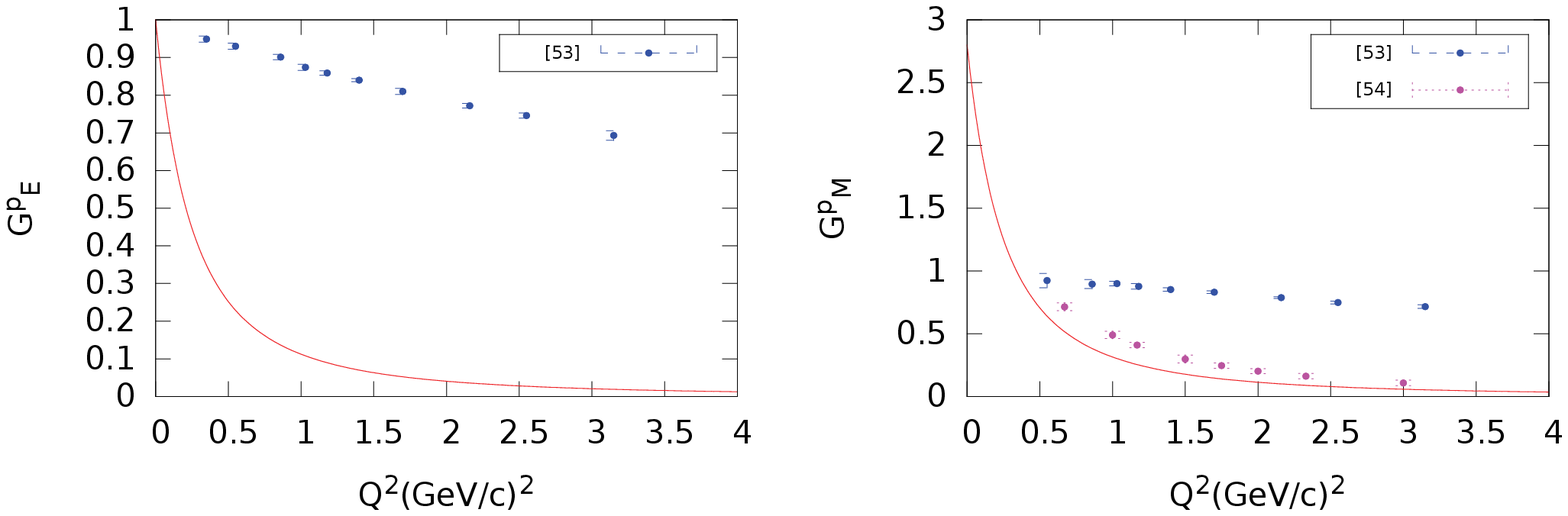}
\caption{(color online). The electric ($ G^{p}_{E}(Q^2)$) and magnetic ($G^{p}_{M}(Q^2)$) form factors of the proton as a function of $Q^2$\,(GeV/c)$^2$. The data has been taken from the references mentioned in the legend.}
\label{fig1-GEp-GMp}
\end{figure}

\begin{figure}
\includegraphics [width=1.\textwidth] {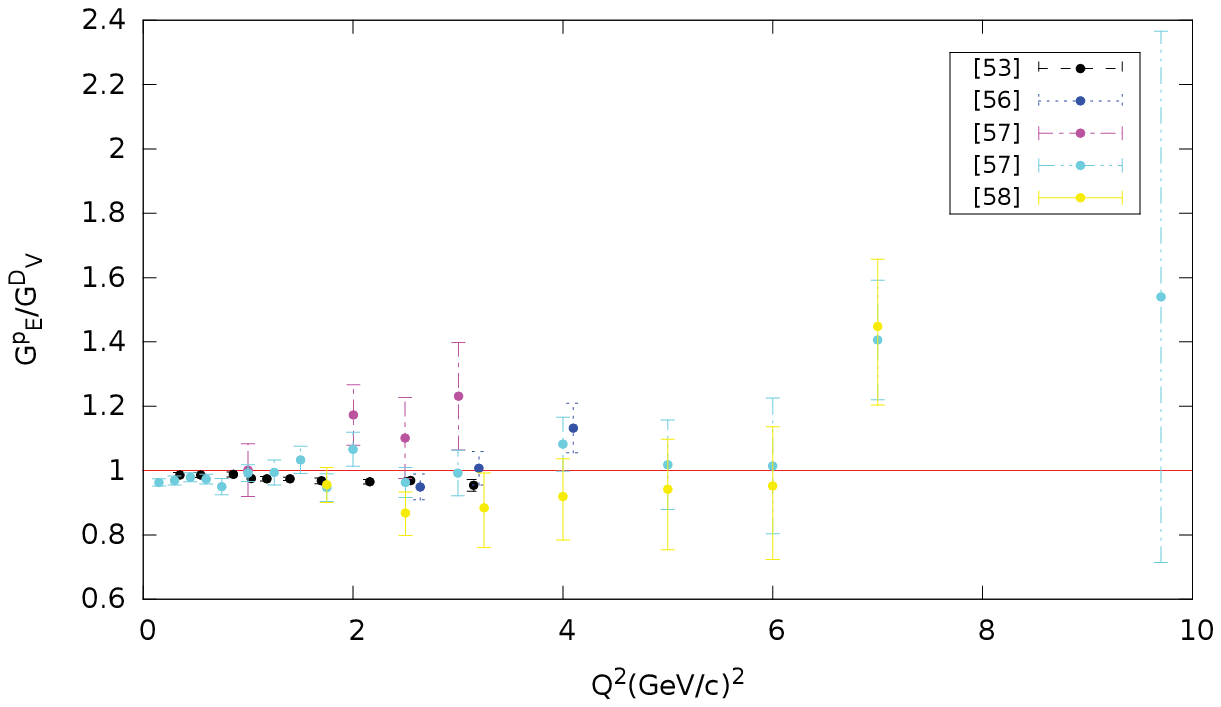}
\caption{(color online). The ratio $ G^{p}_{E}(Q^2)/G^D_V(Q^2)$  as a function of $Q^2$\,(GeV/c)$^2$. The data has been taken from the references mentioned in the legend.}
\label{fig2-GEp-GVD}
\end{figure}

\begin{figure}
\includegraphics [width=1.\textwidth] {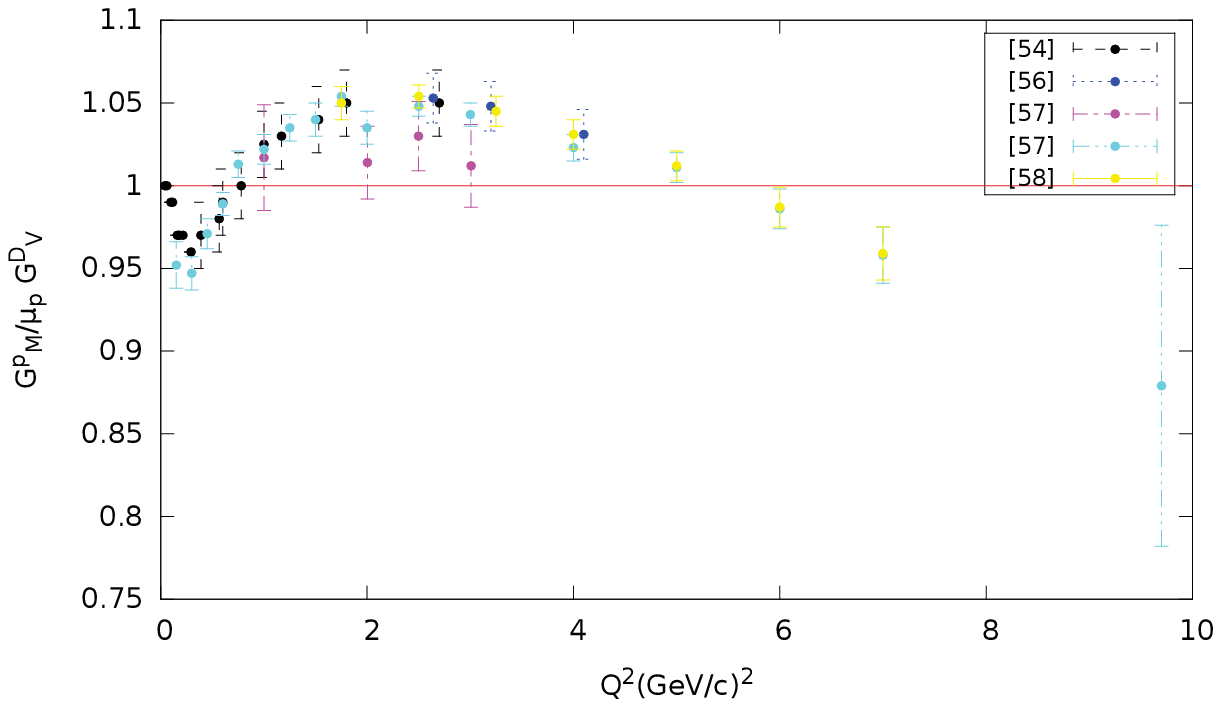}
\caption{(color online). The  ratio  $ G^{p}_{M}(Q^2)/\mu_p G^D_V(Q^2)$   as a function of $Q^2$\,(GeV/c)$^2$. The data has been taken from the references mentioned in the legend.}
\label{fig3-GMp-mupGVD}
\end{figure}

\begin{figure}
\includegraphics [width=1.\textwidth] {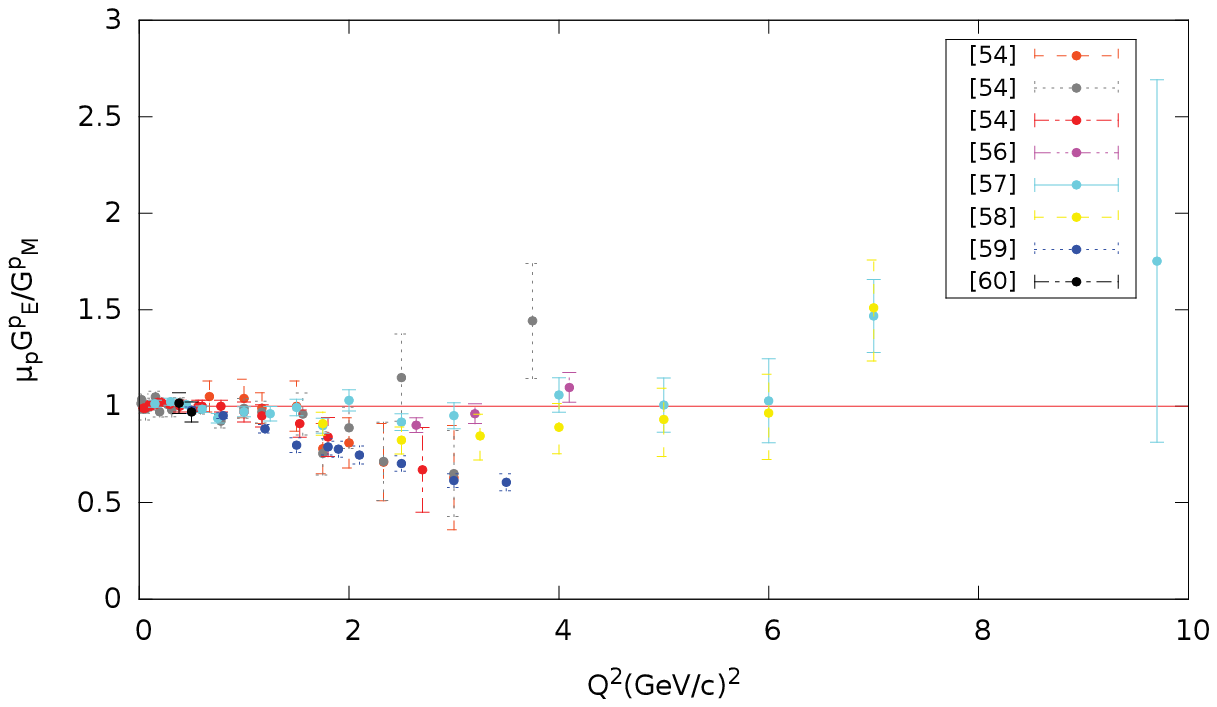}
\caption{(color online). The ratio $ \mu_p G^{p}_{E}(Q^2)/G^{p}_{M}(Q^2)$  as a function of $Q^2$\,(GeV/c)$^2$. The data has been taken from the references mentioned in the legend.}
\label{fig4-mupGEp-GMp}
\end{figure}

In Fig. \ref{fig1-GEp-GMp}, we have presented the variation of electric ($ G^{p}_{E}(Q^2)$) and magnetic ($G^{p}_{M}(Q^2)$) form factors of the proton with $Q^2$. We find that the charge of $p$ ($G^{p}_{E}(Q^2=0)$) is 1 and as the $Q^2$ value increases, it falls off very quickly at small values of till $Q^2\approx 1$\,GeV$^2$. The data however falls off steadily from $\approx 1$ at $Q^2\approx 0$\,GeV$^2$ to $\approx 0.7$ at $Q^2\approx 3$\,GeV$^2$. For the case of $G^{p}_{M}(Q^2)$, the $\chi$CQM results agree quite well with the data points for $Q^2 > 0.5$\,GeV$^2 $ and $Q^2 < 3$\,GeV$^2 $. In the absence of data for $Q^2 < 0.5$\,GeV$^2 $ and $Q^2 > 3$\,GeV$^2 $, it is difficult to compare the results at these values. The magnetic moment of proton $\mu_p= G^{p}_{M}(Q^2=0)$ comes out to be $2.80 \mu_N$ which is in fair agreement with data \cite{PDG}. The ratios $ G^{p}_{E}(Q^2)/G^D_V(Q^2)$, $ G^{p}_{M}(Q^2)/\mu_p G^D_V(Q^2)$ and $ \mu_p G^{p}_{E}(Q^2)/G^{p}_{M}(Q^2)$ have been respectively presented in Fig. \ref{fig2-GEp-GVD}, Fig. \ref{fig3-GMp-mupGVD} and Fig. \ref{fig4-mupGEp-GMp}. The results for $ G^{p}_{E}(Q^2)/G^D_V(Q^2)$ are more or less in agreement with the data. Different data shows the value of $ G^{p}_{E}(Q^2)/G^D_V(Q^2)$ close to 1. Similarly,  for the case of  $ G^{p}_{M}(Q^2)/\mu_p G^D_V(Q^2)$ and $ \mu_p G^{p}_{E}(Q^2)/G^{p}_{M}(Q^2)$, a fair agreement with data is obtained. Even though the data varies from $0.95 - 1.05$, it stays close to $1$. More data for $Q^2 > 4$\,GeV$^2$ may be needed so see if there is some variation from $1$. The proton form factors and their ratios have been measured in the polarization experiments, recoil polarization experiments and beam-target asymmetry measurements  \cite{CEBAF-proton1,CEBAF-proton2,proton-data1,proton-data2,proton-data3,proton-data4,MIT-RSS-proton1,MIT-RSS-proton2,puckett-proton1,puckett-proton2}.

\begin{figure}
\includegraphics [width=1.\textwidth] {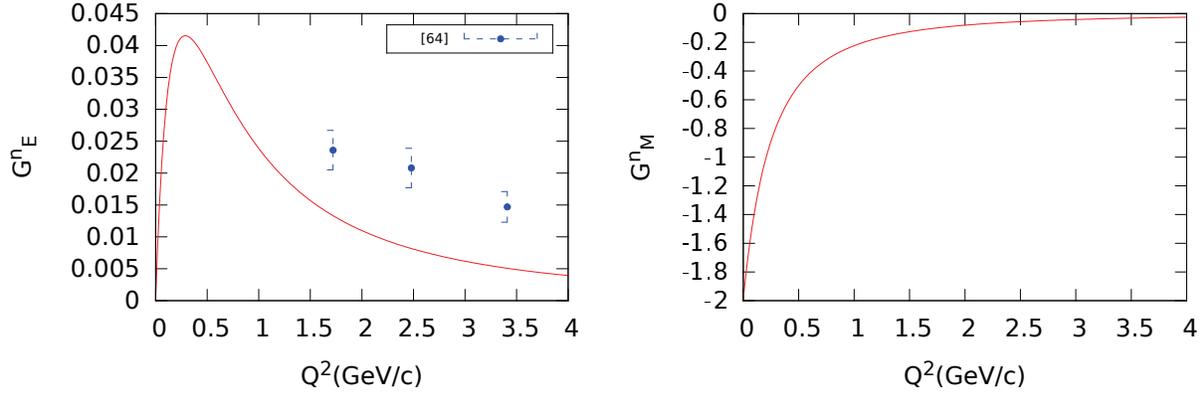}
\caption{(color online). The electric ($ G^{n}_{E}(Q^2)$) and magnetic ($G^{n}_{M}(Q^2)$) form factors of the neutron as a function of $Q^2$\,(GeV/c)$^2$. The data has been taken from the reference mentioned in the legend.}
\label{fig5-GEn-GMn}
\end{figure}

\begin{figure}
\includegraphics [width=1.\textwidth] {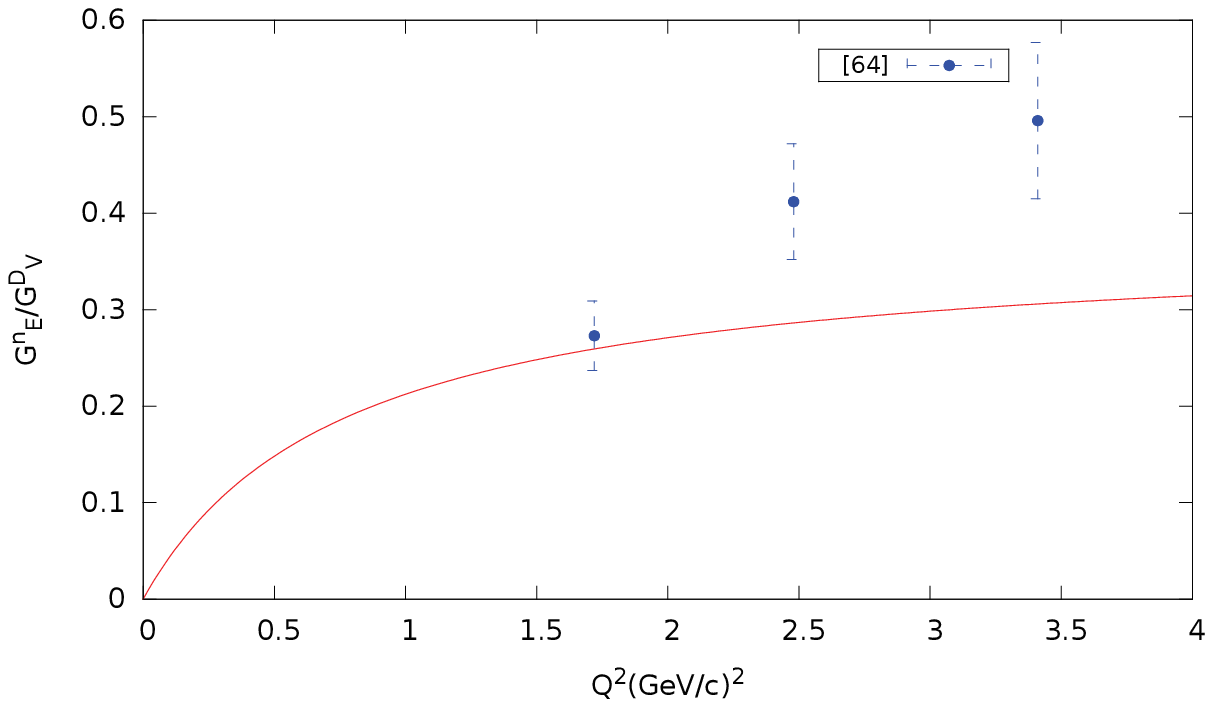}
\caption{(color online). The ratio $ G^{n}_{E}(Q^2)/G^D_V(Q^2)$  as a function of $Q^2$\,(GeV/c)$^2$. The data has been taken from the reference mentioned in the legend.} 
\label{fig6-GEn-GVD}
\end{figure}

\begin{figure}
\includegraphics [width=1.\textwidth] {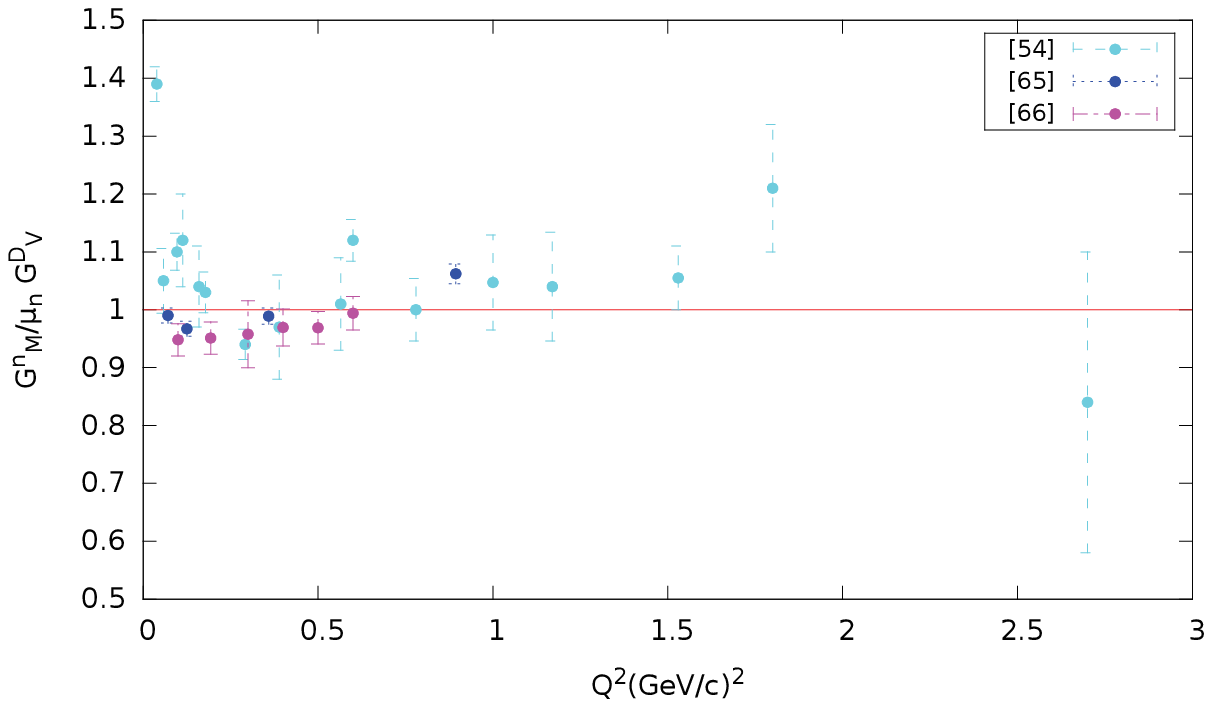}
\caption{(color online). The  ratio  $ G^{n}_{M}(Q^2)/\mu_n G^D_V(Q^2)$   as a function of $Q^2$\,(GeV/c)$^2$. The data has been taken from the references mentioned in the legend. }
\label{fig7-GMn-munGVD}
\end{figure}

\begin{figure}
\includegraphics [width=1.\textwidth] {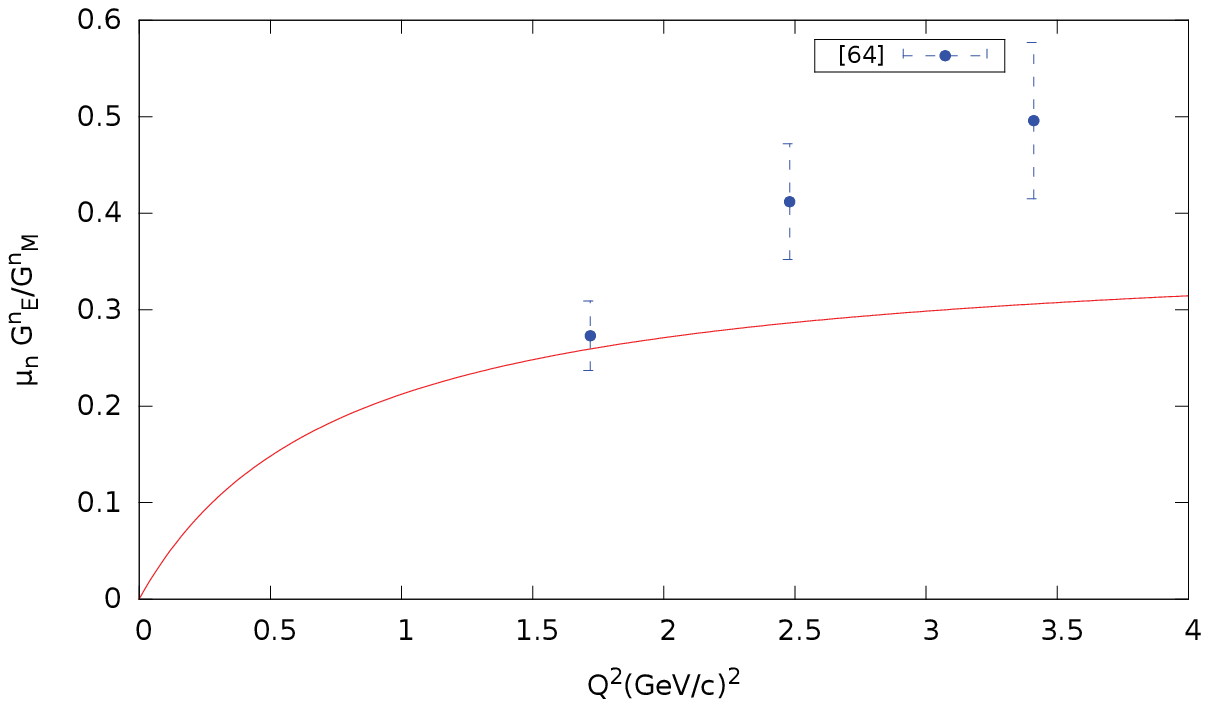}
\caption{(color online). The ratio $ \mu_n G^{n}_{E}(Q^2)/G^{n}_{M}(Q^2)$  as a function of $Q^2$\,(GeV/c)$^2$. The data has been taken from the reference mentioned in the legend.}
\label{fig8-munGEn-GMn}
\end{figure}

\begin{figure}
\includegraphics [width=1.\textwidth] {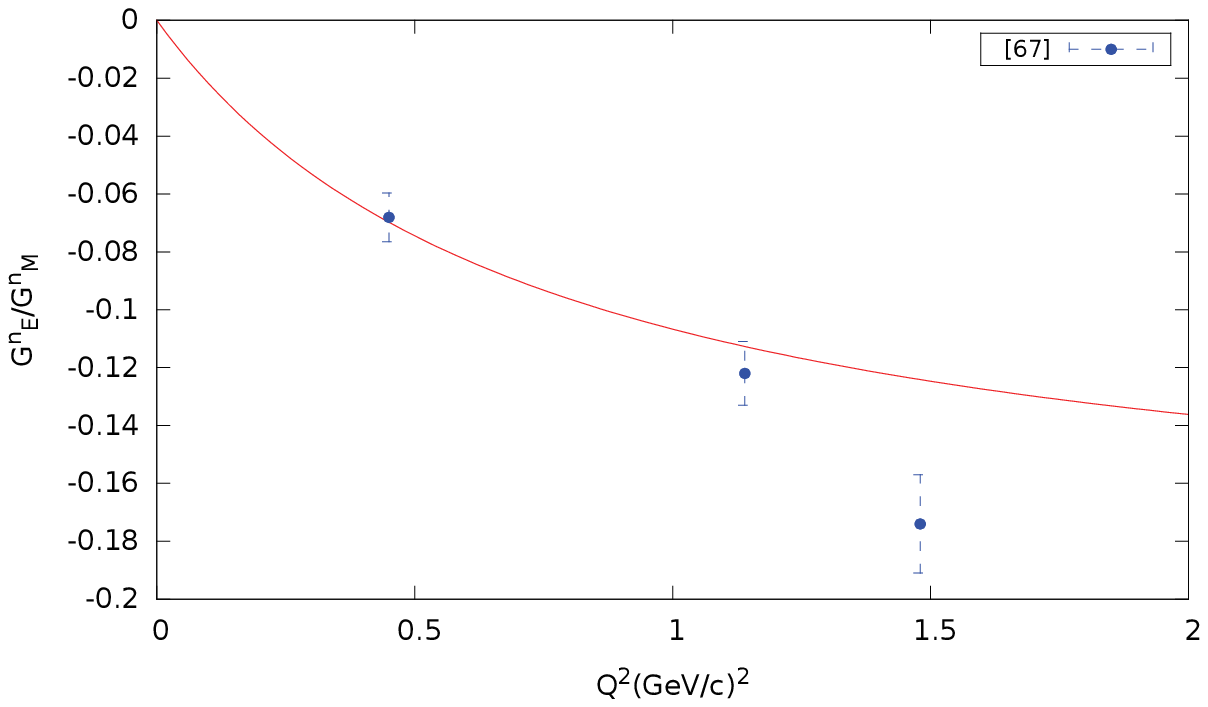}
\caption{(color online). The ratio $ G^{n}_{E}(Q^2)/G^{n}_{M}(Q^2)$  as a function of $Q^2$\,(GeV/c)$^2$. The data has been taken from the reference mentioned in the legend.}
\label{fig9-GEn-GMn}
\end{figure}

In Fig. \ref{fig5-GEn-GMn}, we have presented the variation of electric ($ G^{n}_{E}(Q^2)$) and magnetic ($G^{n}_{M}(Q^2)$) form factors of the neutron with $Q^2$. The ratios $ G^{n}_{E}(Q^2)/G^D_V(Q^2)$, $ G^{n}_{M}(Q^2)/\mu_n G^D_V(Q^2)$ and $ \mu_n G^{n}_{E}(Q^2)/G^{n}_{M}(Q^2)$ have been respectively presented in Fig. \ref{fig6-GEn-GVD}, Fig. \ref{fig7-GMn-munGVD} and Fig. \ref{fig8-munGEn-GMn}. In Fig. \ref{fig9-GEn-GMn}, we have presented the ratio
$ G^{n}_{E}(Q^2)/G^{n}_{M}(Q^2)$.  The neutron form factors have been measured in a series of experiments \cite{kubon,anderson} and our results are in fair agreement with the available experimental data. More data in needed for the profound understanding of the form factors of the neutron.

\begin{figure}
\includegraphics [width=1.\textwidth] {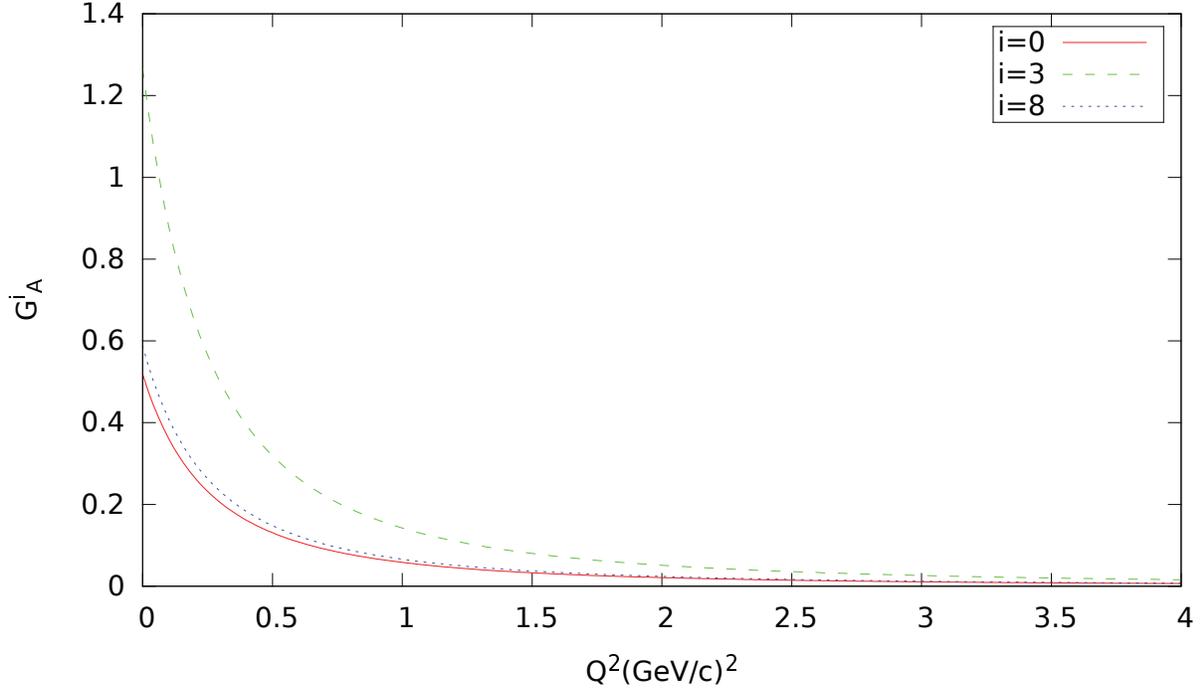}
\caption{(color online). The axial-vector form factors $G^{0}_{A}(Q^2)$, $G^{3}_{A}(Q^2)$ and $G^{8}_{A}(Q^2)$  as a function of $Q^2$\,(GeV/c)$^2$. }
\label{fig10-GAi}
\end{figure}

In Fig. \ref{fig10-GAi}, we have plotted the axial-vector form factors $G^{i}_{A}(Q^2)$ as a function of $Q^2$ for $i=0,3,8$. These axial-vector form factors at $Q^2=0$ give the axial-vector coupling constants: $g_A^0$ corresponds to the flavor singlet component, $g_A^3$ and  $g_A^8$ correspond to the flavor non-singlet components. The present experimental situation for the case of $g_A^0$, $g_A^3$ and  $g_A^8$,
is summarized as follows \cite{PDG}:
\bea
g^{0~\text {expt}}_{A} &=& 0.30\pm 0.06,\nonumber \\
g^{3~\text {expt}}_{A}&=& 1.267 \pm 0.0025, \nonumber \\
g^{8~\text {expt}}_{A}&=& 0.588 \pm 0.033 \label{d8}\,,
\eea whereas the $\chi$CQM results are given as
\bea
g^{0}_{A}&=& 0.519,\nonumber \\
g^{3}_{A} &=& 1.266, \nonumber \\
g^{8}_{A} &=& 0.588\,.
\eea
The $Q^2$ dependence of the singlet and non-singlet form factors varies as
\be G^3_{A}(Q^2)< G^8_{A}(Q^2) < G^0_{A}(Q^2).
\ee

\begin{figure}
\includegraphics [width=1.\textwidth] {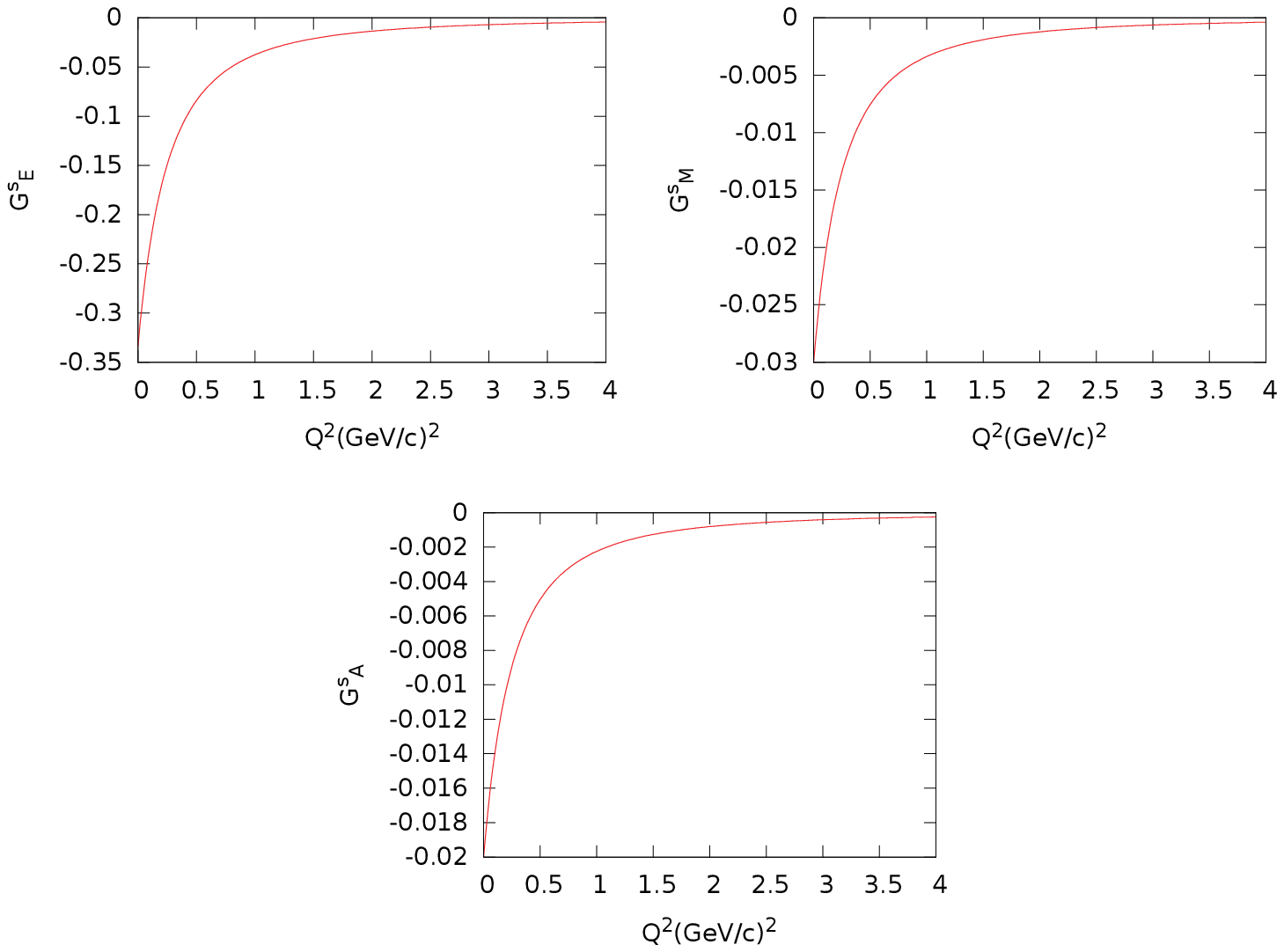}
\caption{(color online). The strangeness form factors $ G^{s}_{E}(Q^2)$, $ G^{s}_{M}(Q^2)$ and $ G^{s}_{A}(Q^2)$  as a function of $Q^2$\,(GeV/c)$^2$.}
\label{fig11-GEMAs}
\end{figure}
The strangeness quark contributions to the vector and axial-vector form factors ($ G^{s}_{E}(Q^2)$, $ G^{s}_{M}(Q^2)$ and $ G^{s}_{A}(Q^2)$)  have been shown in Fig. \ref{fig11-GEMAs} as a function of $Q^2$. It is well known that the strange quarks contribute to the internal properties of the nucleon  because of the presence of the non-constituent ``quark sea'' (Eq. (\ref{basic})) which includes the effects of chiral symmetry breaking
as well as SU(3) symmetry breaking. From Fig. \ref{fig11-GEMAs} we find that the magnitude of $ G^{s}_{E}(Q^2)$, $ G^{s}_{M}(Q^2)$ and $ G^{s}_{A}(Q^2)$ fall off with the increasing value of $Q^2$. The explicit strangeness contribution to the magnetic form factor and the axial-vector form factor is very small as compared to the electric form factor. This is in agreement with the small but significant contribution of strangeness in the nucleon as indicated by SAMPLE at MIT-Bates \cite{sample}, G0 at JLab \cite{g01,g02},
PVA4 at MAMI \cite{a41,a42} and HAPPEX  at JLab \cite{happex1,happex2,happex3,happex4}. A determination of $G_A^s$ at low values of $Q^2$ would have important implications in the determination of strange spin polarization $\Delta s$ which is otherwise zero in the case of nucleon.

\section{summary and conclusions}

To summarize, the  electromagnetic and axial-vector form factors of the nucleon ($G^{p,n}_{E,M}(Q^2)$ and $G^{p,n}_{A}(Q^2)$) have been phenomenologically determined  in the chiral constituent quark model ($\chi$CQM) using the spin observables. The  $\chi$CQM helps in the understanding the dynamics of the constituents of the nucleon affected by chiral symmetry breaking in terms of the  quark flavor contributions to the form factors of the nucleon. Further, in light of the precision data available for increased $Q^2$ range as well as to present a comprehensive analysis of the vector and axial-vector form factors, the calculations have been extended to analyse the $Q^2$ dependence of these quantities using the conventional dipole form of parametrization. The contributions of the quark flavor to the electromagnetic structure of the nucleon have been calculated by combining the  electromagnetic and neutral weak vector currents as well as axial current leading to the flavor decomposition of the form factors ($G^{q}_{E}(Q^2)$, $G^{q}_{M}(Q^2)$ and $G^{q}_{A}(Q^2)$). The contribution of strange quarks provides an ideal probe for the virtual sea quarks present in the nucleon particularly the strange spin polarization  $\Delta s$ which corresponds to the value of the strange axial form factor $G_A^s$ at zero-momentum transfer ($Q^2 = 0$).  Despite considerable efforts in the past few years, the experimental data on $\Delta s$ and $G_A^s$ point out the need for additional refined data. Moreover, the $Q^2$ dependence of $G_A^s$ is also unknown. In the scarcity of precise data at higher $Q^2$ and very low $Q^2$, the  results have been compared with the recent available experimental observations.

In conclusion, we would like to state that our results provide important constraints on the future experiments to describe the explicit role of constituent and non-constituent degrees of freedom particularly the strangeness contribution. Different experiments are contemplating the possibility of performing the high precision measurements over a wide $Q^2$ region in the near future which will help in the profound understanding of the nonperturbative properties of QCD.

\section*{ACKNOWLEDGMENTS}

H.D. would like to thank Department of Science and Technology (Ref No. SB/S2/HEP-004/2013), Government of India, for financial support.

\end{document}